\documentclass[11pt]{article}
\usepackage{moriond,epsfig}
\usepackage{amsmath,amssymb}

\def\be{\begin{equation}}
\def\ee{\end{equation}}
\def\bea{\begin{eqnarray}}
\def\eea{\end{eqnarray}}

\begin{document}
\vspace*{4cm}
\title{SEARCHES OF STELLAR MASS DARK MATTER FROM AN ANALYSIS OF VARIABILITIES OF HIGH RED-SHIFTED QSOs}

\author{ A.F. ZAKHAROV,$^{1,2}$ L.\v C. POPOVI\'C,$^{3,4}$ P. JOVANOVI\'C$^3$}

\address{$^1$Institute of Theoretical and
Experimental Physics,\\
           25, B.Cheremushkinskaya st., Moscow, 117259, Russia\\
  $^2$Astro Space
Centre of Lebedev Physics Institute, Moscow\\
 $^3$Astronomical Observatory of Beograd,\\
  Volgina 7, 11160 Beograd, Serbia \\
  $^4$Astrophysikalisches Institut Potsdam, \\
  An der Sternwarte 16, 14482 Potsdam, Germany\\}

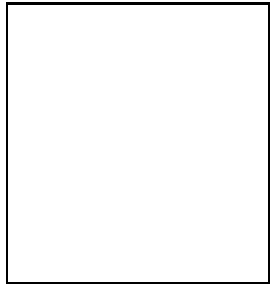
\maketitle\abstracts{ We consider a contribution of microlensing
to the X-ray variability of high-redshifted QSOs. Such an effect
could be caused by stellar mass objects (SMO) located in a bulge
or/and in a halo of this quasar as well as  at cosmological
distances between an observer and a quasar.
%Here, we  not consider
%microlensing caused by deflectors in our Galaxy  since it is
%well-known from recent MACHO, EROS and  OGLE observations that the
%corresponding optical depth for the Galactic halo and the Galactic
%bulge is lower than $10^{-6}$.
Cosmologically distributed
gravitational microlenses could be localized in galaxies (or even
in bulge or halo of  gravitational macrolenses) or could be
distributed in a uniform way. We have analyzed both cases of such
distributions. As a result of our analysis, we obtained that the
optical depth for microlensing caused by stellar mass objects is
usually small for quasar bulge and  quasar halo gravitational
microlens distributions ($\tau\sim 10^{-4}$). On the other hand,
the optical depth for gravitational microlensing caused by
cosmologically distributed deflectors could be significant and
could reach $10^{-2} - 0.1$ at $z\sim 2$.  This means that
cosmologically distributed deflectors may contribute
significantlly  to the X-ray variability of high-redshifted QSOs
($z>2$). Considering that the upper limit of the optical depth
($\tau\sim 0.1$) corresponds to the case where dark matter forms
cosmologically distributed deflectors,   observations of the X-ray
variations of unlensed QSOs can be used for the estimation of the
dark matter fraction of microlenses. }

%\section{Introduction}

The X-ray radiation of Active Galactic Nuclei (AGNs), in the
continuum as well as in spectral lines, has rapid and irregular
variability.\cite{Man02}
 X-ray flux variability has long been known to be a common property of
active galactic nuclei (AGNs), e.g. Ariel 5 and HEAO 1 first
revealed long-term (days to years) variability in AGNs  and by
uninterrupted observations of EXOSAT  rapid (thousands of seconds)
variability was also established as common in these sources (see,
for example reviews \cite{Mush93,Ulrich97} and references
therein). X-ray flux variations are observed on timescales from
$\sim$1000 s to years, and amplitude variations of up to an order
of magnitude are observed in the $\sim$ 0.1 - 10 keV band.
Recently, Manners at al.\cite{Man02}
 analyzed the variability of a sample of 156 radio-quiet quasars
taken from the ROSAT archive, considering the trends in
variability of the amplitude with luminosity and with redshift.
They found that there were evidences for a growth in AGN X-ray
variability amplitude towards high redshift ($z$) in the sense
that AGNs of the same X-ray luminosity were more variable at
$z>2$. They explained the $\sigma$ {\it vs.} $z$ trend assuming
that the high-redshifted AGNs accreted at a larger fraction of the
Eddington limit than the low-redshifted ones.

On the other hand, the contribution of microlensing to AGN
variability was considered in many papers (see e.g.
papers\cite{Hawk93,Hawk02,Wamb01,Wamb01b,Zakh97}, and references
therein). Moreover, recently X-ray microlensing of AGN has been
considered.\cite{Popov01a,Pop01b,Popovic03,Tak01,Chart02a,Dai03}
Taking into account that the X-rays of AGNs are generated in the
innermost and very compact region of an accretion disc, the X-ray
radiation in the continuum as well as in a line can be strongly
affected by microlensing.\cite{Popovic03} Simulations of X-ray
line profiles are presented in a number of papers, see, for
example,
papers\cite{Zak_rep02,Zak_rep02a,Zak_rep02_xeus,Zak_rep03_ASR,Zak_rep03_Su}
and references therein, in particular Zakharov et al.\cite{ZKLR02}
showed that an information about magnetic field may be extracted
from X-ray line shape analysis; Zakharov \&
Repin\cite{Zak_rep03_AA} discussed signatures of X-ray line shapes
for highly inclined accretion disks.

 Recent
observations of three lens systems seem to support this
idea.\cite{Osh01,Chart02a,Dai03}  Popovi\'c et al.\cite{Popovic03}
showed that objects in a foreground galaxy with very small masses
can cause strong changes in the X-ray line profile.  This  fact
may indicate that the observational probability of X-ray variation
due to microlensing events is higher  than in the UV and optical
radiation of AGNs. It is connected with the fact that typical
sizes of X-ray emission regions are much smaller than typical
sizes of those producing optical and UV bands. Typical optical and
UV emission region sizes could be comparable or even larger than
Einstein radii of microlenses and therefore microlenses magnify
only a small part of the region emitting in the optical or UV band
(see e.g. papers,\cite{Pop01b,Aba02} for UV and optical spectral
line region). This is reason that it could be a very tiny effect
from an observer point of view.

The aim of the work is to discuss the contribution of microlensing
to the relation $\sigma$ {\it vs.} $z$ for X-ray radiation
considering the recent results given  in
papers.\cite{Man02,Popovic03} The results of calculations and
detailed discussions are presented in the paper.\cite{Zakharov04}

Just after the discovery of the first multiple-imaged quasar
QSO~0957+561~A,B by Walsh et al.\cite{Walsh79} the idea of
microlensing by low mass stars in a heavy halo was suggested by
Gott.\cite{Gott81} First evidence of quasar microlensing was found
by Irwin et al.\cite{Irwin89} Now there is a number of known
gravitational lens systems\cite{Claeskens02,Browne03} and some of
them show evidence for microlensing.\cite{Wamb01}

More than 10 years ago Hawkins\cite{Hawk93} (see also
paper\cite{Hawk02})
 put
forward the idea that nearly all quasars are being microlensed.
Hawkins\cite{Hawk02} argued that the observational results  favor
the disc instability model for Seyfert galaxies, and the
microlensing model for quasars. The starburst and disc instability
models are ruled out for quasars, while the microlensing model is
in good agreement with the observations.

As  was mentioned earlier by Popovi\'c et al.\cite{Popovic03} the
probability of microlensing by stars or other compact objects in
halos and bulges of quasars is very low (about $10^{-4} -
10^{-3}$). However, for cosmologically distributed microlenses it
could reach $10^{-2} - 0.1$ at $z\sim 2$. The upper limit $\tau
\sim 0.1$ corresponds to the case where compact dark matter forms
cosmologically distributed microlenses.
 This indicates that   such a phenomenon could be  observed
frequently, but only for distant sources ($z \sim 2$). Moreover,
it is in good agreement with the trend in the variability
amplitude with redshift,\cite{Man02} where AGNs of the same X-ray
luminosity are more variable at $z>2$.

To investigate  distortions of spectral line shapes due to
microlensing the most real candidates are multiply imaged quasars.
However, for these cases the simple point-like microlens model may
not be very good approximation\cite{Wamb01,Wamb01b} and one should
use a numerical approach, such as the MICROLENS ray tracing
program, developed by J. Wambsganss  or  some analytical approach
for magnification near caustic curves like
folds\cite{Schneider92a} or near singular caustic points like
cusps\cite{Schneider92,Zakharov95,Zakharov97,Zakharov99} as was
realized by Yonehara.\cite{Yonehara01}

If  we believe in the  observational arguments\cite{Hawk02} that
the variability of a significant fraction of distant quasars is
caused by microlensing,  the analysis of the properties of X-ray
line shapes due to microlensing\cite{Popovic03} is a powerful tool
to confirm or rule out the Hawkins hypothesis.

As it was mentioned, the probability that the shape of the  Fe
$K\alpha$ line is distorted (or amplified) is highest in
gravitationally lensed systems.
 Actually, this phenomena was
discovered recently\cite{Osh01,Dai03,Chart02a,Cha02b,Cha04} who
found evidence for such an effect for   QSO H1413+117 (the
Cloverleaf, $z=2.56$), QSO~2237+0305 (the Einstein Cross,
$z=1.695$), MG J0414+0534 ($z=2.64$) and possibly for BAL
QSO~08279+5255 ($ z=3.91$).
 One could say that it is
natural that the discovery of X-ray microlensing was made for this
quasar, since the Einstein Cross QSO~2237+0305 is the most
"popular" object to search for microlensing, because the first
cosmological microlensing  phenomenon was found in this
object\cite{Irwin89} and  several groups have been monitoring the
quasar QSO~2237+0305 to find evidence for microlensing.
Microlensing has been suggested for the quasar MG
J0414+0534\cite{Angonin99} and for the quasar QSO
H1413+117.\cite{Chae01} Therefore, in future may be a chance to
find X-ray microlensing for other gravitationally lensed systems
that have  signatures of microlensing in the optical and radio
bands. Moreover, considering the sizes of the sources of X-ray
radiation, the variability in the X-ray range during microlensing
event should be more prominent than in the optical and UV.
Consequently, gravitational microlensing in the X-ray band
 is a powerful tool for
 dark matter investigations, as the upper limit of optical
depth ($\tau\sim 0.1$) corresponds to the case where dark matter
forms cosmologically distributed deflectors.
 The observed rate of
 microlensing can be used for estimates of the cosmological density of
 microlenses, but  durations of
 microlensing events could be used to estimate microlens masses.\cite{Wamb01,Wamb01b}

From our calculations we can conclude:\cite{Zakharov04}

i) The optical depth in the bulge and halo of host galaxy is $\sim
10^{-4}$. This is in good agreement with previous
estimates.\cite{Popovic03} Microlensing by  deflectors from the
host galaxy halo and bulge makes a minor contribution to the X-ray
variability of QSOs.

ii) The optical depth for  cosmologically distributed deflectors
could
 be $\sim 10^{-2}-0.1$ at $z\sim 2$ and might contribute  significantly
 to the
X-ray variability of high-redshift QSOs. The value $\tau\sim 0.1$
corresponds to the case where compact dark matter forms
cosmologically distributed microlenses.

iii) The optical depth for  cosmologically distributed deflectors
($\tau_L^p$)  is higher for $z>2$ and increases slowly beyond
$z=2$.
 This indicates that the contribution of microlensing on the
X-ray variability of QSOs with redshift $z>2$ may be significant,
and also  that this contribution could  be nearly constant for
high-redshift QSOs. This is in good agreement with the fact that
AGNs of the same X-ray luminosity are more variable at
$z>2$.\cite{Man02}

 iv) Observations of X-ray variations of unlensed QSOs can be used for estimations of matter fraction of microlenses.
 The rate of
 microlensing can be used for estimates of the cosmological density of
microlenses, and consequently the fraction of dark matter
microlenses, but the durations of microlensing events could be
used for gravitational microlens mass estimations.

\section*{Acknowledgements}

This work was supported in part by  the Ministry of Science,
Technologies and Development of Serbia through the project
"Astrophysical Spectroscopy of Extragalactic Objects" (L\v CP \&
PJ) and the Alexander von Humboldt Foundation through the program
for foreign scholars (L\v CP).

AFZ is grateful to the organizers of the XXXIXth Rencontres de
Moriond Conference "Exploring the Universe: Contents and
structures of the Universe"  for the hospitality in La Thuile.


\section*{References}
\begin{thebibliography}{}

\bibitem{Man02}
J. Manners, O. Almaini \& A. Lawrence,   {\it MNRAS}, {\bf 330},
390 (2002).

\bibitem{Mush93}
R.F. Mushotzky, C. Done \& K.A. Pounds,  {\it ARA\&A}, {\bf 31},
717 (1993).

\bibitem{Ulrich97}
M.-H. Ulrich, L. Maraschi  \& C.M. Megan, {\it  ARA\&A}, {\bf 35},
445 (1997).

\bibitem{Hawk93}
 M.R.S. Hawkins, {\it  Nature},  {\bf 366}, 242 (1993).

\bibitem{Hawk02}
 M.R.S. Hawkins, {\it  MNRAS},  329, 76 (2002).

\bibitem{Wamb01}
J.   Wambsganss,
 in {\it Microlensing 2000: A new Era of Microlensing
 Astrophysics,}
ed. J.W.Menzies and P.D.Sackett
 ASP Conf. Series, 239, 351 (2001).

\bibitem{Wamb01b}
J.  Wambsganss,   {\it PASA}, {\bf 18}, 207 (2001).

\bibitem{Zakh97}
A.F. Zakharov, {\it Gravitational lenses and microlensing,}
(Janus-K, Moscow, 1997).

\bibitem{Popov01a}
 L. \v C. Popovi{\'c}, E.G.  Mediavilla, J. Mu\~noz  et al.
{\it Serb. Aston. J.}, 164, 73 (2001).
%(Also, presented on GLITP Workshop on Gravitational Lens
%Monitoring, 4-6 June 2001, La Laguna, Tenerife, Spain)

\bibitem{Pop01b}
 L. \v C. Popovi{\'c}, E.G.  Mediavilla \& J. Mu\~noz,
{\it A{\&}A} 378, 295 (2001).

\bibitem{Popovic03}
 L. \v C. Popovi\'c, E.G. Mediavilla, P. Jovanovi\'c \& J.A. Mu\~noz,
{\it A \&  A}, 398, 975 (2003).

\bibitem{Tak01}
R. Takahashi, A. Yonehara \& S. Mineshige,  {\it  PASJ}, {\bf 53},
387 (2001).

\bibitem{Chart02a}
G. Chartas, E. Agol, M. Eracleous et al., {\it ApJ}, {\bf 568},
509 (2002).

\bibitem{Dai03}
X. Dai, G. Chartas, E. Agol et al., {\it ApJ}, {\bf 589}, 100
(2003).

\bibitem{Zak_rep02}
A.F.   Zakharov \& S.V. Repin,  {\it Astronomy Reports}, {\bf 46},
360 (2002).

\bibitem{Zak_rep02a}
A.F.   Zakharov \& S.V. Repin,
 in {\it Proc. of the Eleven Workshop
   on General Relativity  and Gravitation in Japan,} ed.
   J.~Koga, T.~Nakamura, K.~Maeda, K.~Tomita, (Waseda University,
   Tokyo, 2002) 68.

\bibitem{Zak_rep02_xeus}
A.F.   Zakharov \& S.V. Repin, in {\it   Proc. of the Workshop
"XEUS - studying the evolution of the hot
   Universe"}, ed. G. Hasinger et al.,  MPE Report~281,
   339 (2002).

\bibitem{Zak_rep03_ASR}
A.F.   Zakharov \& S.V. Repin, {\it Advances in Space Res.}
(accepted).

\bibitem{Zak_Ma_New_As04}
A.F. Zakharov, Z. Ma, Y. Bao,   {\it New Astronomy} (accepted).

\bibitem{Zak_rep03_Su}
A.F.   Zakharov \& S.V. Repin, in {\it   Proc. of the 214th IAU
Symposium "High Energy Processes and Phenomena in
   Astrophysics"}, ed. X.D. Li, V. Trimble, Z.R. Wang,  {\bf 214},
   97 (2003).

\bibitem{ZKLR02}
A.F.   Zakharov, N.S.  Kardashev, V.N. Lukash  \& S.V. Repin, {\it
MNRAS}, {\bf 342}, 1325 (2003).

\bibitem{Zak_rep03_AA}
A.F.  Zakharov \& S.V. Repin, {\it A\& A}, {\bf 406}, 7 (2003).

\bibitem{Osh01}
T. Oshima, K. Mitsuda, R. Fujimoto et al., {\it ApJ}, {\bf
563}, L103 (2001).

\bibitem{Aba02}
C. Abajas, E.G. Mediavilla, J.A. Mu\~noz et al., {\it ApJ}, {\bf
576}, 640 (2002).

\bibitem{Zakharov04}
A.F. Zakharov, L.\v C. Popovi\'c \& P. Jovanovi\'c,  {\it A \& A},
{\bf 420}, 881 (2004).
%(accepted); astro-ph/0403254.

\bibitem{Walsh79}
D. Walsh, R.F. Carswell \& R.J. Weymann, {\it Nature}, {\bf 279},
381 (1979).

\bibitem{Gott81}
J.R. Gott, {\it ApJ}, {\bf 243}, 140 (1981).

\bibitem{Irwin89}
M.J. Irwin, R.L. Webster, P.C. Hewett et al.,
% Corrigan R.T., Jedrzejewski R.I.
{\it AJ}, {\bf 98}, 1989 (1989).

\bibitem{Claeskens02}
J.F.Claeskens \& J. Surdej, {\it Astron \& Astroph. Rev.}, {\bf
10}, 263 (2002).

\bibitem{Browne03}
I.W.A. Browne, P.N. Wilkinson, N.J.F. Jackson et al., {\it MNRAS},
{\bf 341}, 13 (2003).


\bibitem{Schneider92a}
P. Schneider, J. Ehlers \& E.E. Falco,  {\it Gravitational
Lenses,} (Springer, Berlin, 1992).


\bibitem{Schneider92}
P. Schneider \& A. Weiss, {\it A \& A}, {\bf 260}, 1 (1992).



\bibitem{Zakharov95}
A.F. Zakharov,   {\it A \&  A}, {\bf 293}, 1 (1995).


\bibitem{Zakharov97}
A.F. Zakharov,  {\it Ap\&SS}, {\bf 252}, 369 (1997).

\bibitem{Zakharov99}
A.F. Zakharov,  in  {\it Recent Developments in Theoretical and
Experimental General Relativity, Gravitation, and Relativistic
Field Theories},  ed. Ts.~Piran and R.~Ruffini (World Scientific
Publishers, Singapore, 1999), 1500.

\bibitem{Yonehara01}
A.   Yonehara, {\it  PASA}, {\bf 18}, 211 (2001).

\bibitem{Cha02b}
G. Chartas, W.N. Brandt, S.C. Gallagher \& G.P. Garmire, {\it
ApJ}, {\bf 579}, 169 (2002).

\bibitem{Cha04}
G. Chartas, M. Eracleous, E. Agol \& S.C. Gallagher,   {\it ApJ},
{\bf 606}, 78 (2004).% accepted (astro-ph/0401240).


\bibitem{Angonin99}
M.-C. Angonin-Willaime, C. Vanderriest, F. Courbin et al.,
 % Burud, I.; Magain, P.; Rigaut, F.
%About the origin of extinction in the gravitational lens system MG J0414+0534
{\it A\&A}, {\bf 347}, 434 (1999).



\bibitem{Chae01}
K.-H. Chae, D.A. Turnshek, R.E. Schulte-Ladbeck et al.,
 {\it ApJ},  568, 509 (2001).


\end{thebibliography}
\end{document}